\documentclass[12pt,tightenlines]{revtex4}
\usepackage{amsfonts}
\usepackage{amsmath}
\usepackage{txfonts}
\usepackage{graphicx}
\usepackage{amssymb}
\usepackage{color}
\usepackage{wrapfig,lipsum,booktabs}
\usepackage{floatrow,float}
\usepackage{epstopdf}
\usepackage{hyperref}
\usepackage[normalem]{ulem}

\newcommand{\beginsupplement}{
        \setcounter{table}{0}
        \renewcommand{\thetable}{S\arabic{table}}
        \setcounter{figure}{0}
        \renewcommand{\thefigure}{S\arabic{figure}}
     }
     
\newcommand{\Hgr}{{\sf H}}

\begin{document}
\title{Replays of spatial memories suppress topological fluctuations in cognitive map}
\author{Andrey Babichev$^{1}$, Dmitriy Morozov$^{2}$ and Yuri Dabaghian$^{3}$\textsuperscript{*},}
\affiliation{$^1$Department of Computational and Applied Mathematics, Rice University, Houston, TX 77005, USA \\
$^2$Lawrence Berkeley National Laboratory, Berkeley, CA 94720, USA\\
$^3$Department of Neurology, The University of Texas McGovern Medical School, 6431 Fannin St, Houston, TX 77030\\
$^{*}$e-mail: Yuri.A.Dabaghian@uth.tmc.edu}
\date{\today}

\begin{abstract} 
\vspace{10 mm}

\textbf{Abstract}. 
The spiking activity of the hippocampal place cells plays a key role in producing and sustaining an internalized 
representation of the ambient space---a cognitive map. These cells do not only exhibit location-specific spiking 
during navigation, but also may rapidly replay the navigated routs through endogenous dynamics of the hippocampal 
network. Physiologically, such reactivations are viewed as manifestations of ``memory replays'' that help to learn new 
information and to consolidate previously acquired memories by reinforcing synapses in the parahippocampal networks. 
Below we propose a computational model of these processes that allows assessing the effect of replays on acquiring a 
robust topological map of the environment and demonstrate that replays may play a key role in stabilizing the hippocampal 
representation of space.
\end{abstract}

\maketitle

\newpage

\section{Introduction}
\label{section:intro}

Spatial awareness in mammals is based on an internalized representation of the environment---a cognitive map. 
In rodents, a key role in producing and sustaining this map is played by the hippocampal place cells, which 
preferentially fire action potentials as the animal navigates through specific domains of a given environment---their 
respective place fields. Remarkably, place cells may also activate due to the endogenous activity of the hippocampal 
network during quiescent wake states \cite{Johnson,Pastalkova} or sleep \cite{Wilson,Louie,Ji}. For example, the 
animal can preplay place cell sequences that represent possible future trajectories while pausing at ``choice 
points'' \cite{Papale}, or replay sequences that recapitulate the order in which the place cells have fired during 
previous exploration of the environment \cite{Foster, Hasselmo}. Moreover, spontaneous replays are also observed 
during active navigation, when the hippocampal network is driven both by the idiothetic (body-derived) inputs and 
by the network's autonomous dynamics \cite{Karlsson1,Carr,Jadhav1,Jadhav2,Dragoi}. 

Neurophysiologically, place cell replays are viewed as manifestations of the animal's ``mental explorations'' 
\cite{Zeithamova,Hopfield,Dabaghian,SchemaM}, which help constructing the cognitive maps and consolidating memories 
\cite{Roux,Ego,Girardeau1,Girardeau2,Gerrard}. Although the detailed mechanisms of these phenomena remain 
unknown, it is believed that replays may reinforce synaptic connections that deteriorate over extended periods of 
inactivity \cite{Singer,Sadowski1,Sadowski2}. 

The activity-dependent changes in the hippocampal network's synaptic architecture occur at multiple timescales 
\cite{Karlsson2,Bi,Fusi}. In particular, statistical analyses of the place cells' spiking times indicate that place 
cells exhibiting frequent coactivity tend to form short-lived ``\textit{cell assemblies}''---commonly viewed as 
\textit{functionally} interconnected groups of neurons that form and disband at a timescale between tens of milliseconds 
\cite{Harris,Syntax,Atallah,Bartos} to minutes or longer \cite{Kuhl,Murre,Billeh,Rakic,Hiratani,Russo,Zenke}, i.e., 
that the functional architecture of this network is constantly changing. In \cite{MWind1,MWind2, PLoZ} we used a 
computational model to demonstrate that despite the rapid rewirings, such a ``transient'' network can produce a 
stable topological map of the environment, provided that the connections' decay rate and the parameters of spiking 
activity fall into the physiological range \cite{PLoS,Arai,Basso}. Below we adopt this model to study the role of 
the hippocampal replays in acquiring a robust cognitive map of space. Specifically, we demonstrate that reinforcing 
the cell assemblies by replays helps to reduce instabilities in the large-scale representation of the environment 
and to reinstate the correct topological structure of the cognitive map.

\section{The model}
\label{section:model}

\textbf{1. General description}. The topological model of spatial learning rests on the insight that the hippocampus 
produces a topological representation of spatial environments and of mnemonic memories---a rough-and-ready framework 
that is filled with geometric details by other brain regions \cite{eLife}. This approach, backed up by a growing number 
of experimental \cite{Gothard,Leutgeb,Wills,Touretzky,Moser,Yoganarasimha,Knierim,Fenton,Alvernhe} and computational
\cite{Chen,Petri,Curto} studies, allows using a powerful arsenal of methods from Algebraic Topology, in particular 
Persistent \cite{Carlsson1,Lum,Singh} and Zigzag \cite{Carlsson2,Carlsson3} homology theory techniques, for studying 
structure and dynamics of the hippocampal map. In particular, the approach developed in 
\cite{PLoS,Arai,Basso,Hoffman,CAs,SchemaS} helps to explain how the information provided by the individual place cells 
combines into a large-scale map of the environment, to follow how the topological structure of this map unfolds in time 
and to evaluate the contributions made by different physiological parameters into this process. It was demonstrated, e.g., 
that the ensembles of rapidly recycling cell assemblies can sustain stable qualitative maps of space, provided that the 
network's rewiring rate is not too high. Otherwise the integrity of the cognitive map may be overwhelmed by topological 
fluctuations \cite{MWind1,MWind2,PLoZ}.

\begin{figure}[!h]
	\includegraphics[scale=0.83]{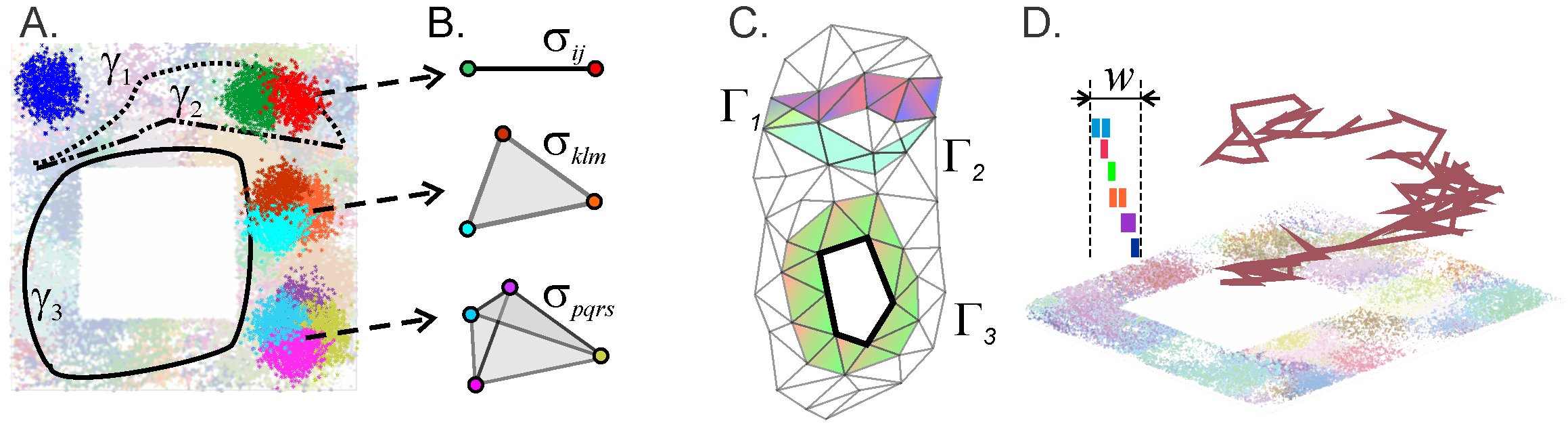}
	\caption{{\footnotesize\textbf{Basic notions of the hippocampal physiology in the context of the topological model.}
			\textbf{A}. In the following, we simulate rat's navigation in a square-shaped environment $\mathcal{E}$ with 
			a hole in the middle. The curves $\gamma_1$, $\gamma_2$ and $\gamma_3$ represent a few short segments of a 
			physical trajectorynavigated by the rat. The place fields---the regions where the corresponding place cells 
			become active---densely cover the environment, forming a place field map $M_{\mathcal{E}}$. An exemplary place 
			field is represented by a highlighted cluster of blue dots in the top left corner of the environment. The right 
			segment of the environment shows a highlighted pair, a triple and a quadruple of the overlapping place fields; 
			the remaining place fields are dimmed into background. 
			\textbf{B}. Simplexes that correspond to overlapping place fields: a single vertex corresponds to place field 
			(or a single active cell); a link between two vertexes represents a pair of overlapping place fields (or a pair 
			of coactive cells); three overlapping place fields (or a triple of coactive place cells) correspond to a triangle, 
			and so forth. 
			\textbf{C}. A collection of simplexes forms a simplicial complex, which schematically represents the net structure 
			of the place field map. Shown is a fragment of a two-dimensional ($2D$) coactivity complex with simplicial paths 
			$\Gamma_1$, $\Gamma_2$ and $\Gamma_3$ that represent the physical paths $\gamma_1$, $\gamma_2$ and $\gamma_3$ 
			shown on the left. The classes of equivalent simplicial paths describe the topological structure of the coactivity 
			complex: the number of topologically inequivalent, contractible simplicial paths such as $\Gamma_1$ and $\Gamma_2$, 
			defines the number of pieces, $b_0$, of the coactivity complex (Section~\ref{section:methods}). The number of 
			topologically inequivalent paths contractible to a one-dimensional ($1D$) loop defines the number $b_1$ of holes 
			and so forth \cite{Hatcher}. 
			\textbf{D}. A schematic representation of a replayed sequence of place cells, shown over the corresponding place 
			fields. The colored ticks in the top left corner schematically represent a sequence of spikes replayed within a	
			short time window $w$.}}
	\label{Figure1}
\end{figure}

Mathematically, the method is based on representing the combinations of coactive place cells in a topological framework, 
as simplexes of a specially designed simplicial complex (Fig.~\ref{Figure1}A,B). Each individual simplex $\sigma$ 
schematically represents a connection (e.g., an overlap) between the place fields encoded by the corresponding place cells' 
coactivity. The full set of such simplexes---the coactivity simplicial complex $\mathcal{T}$---incorporates the entire 
pool of connections encoded by the place cells in a given environment $\mathcal{E}$, and hence represents the topological 
structure of the cognitive map of the navigated space \cite{OKeefe,Best}.

\textbf{2. The topological structure of the coactivity complex}\label{Gam} provides a convenient framework for representing 
spatial information encoded by the place cells. For example, the combinations of the cells ignited during the rat's moves 
along a physical trajectory $\gamma$, or during a mental replay of such a trajectory, is represented by a ``simplicial 
path''---a chain of simplexes $\Gamma = \{\sigma_1, \sigma_2, \ldots, \sigma_k \}$ that qualitatively represents the shape 
of $\gamma$. A simplicial path that loops onto itself represents a closed physical rout; a pair of topologically equivalent 
simplicial paths represent two similar physical paths and so forth (Fig.~\ref{Figure1}C). 

The net structure of the simplicial paths running through a given simplicial complex $\mathcal{T}$ can be used to describe 
its topological shape. Specifically, the number of topologically distinct (counted up to topological equivalence) closed 
paths that contract to zero-dimensional vertexes---the zeroth Betti number $b_0(\mathcal{T})$---enumerates the connected 
components of $\mathcal{T}$; the number of topologically distinct paths that contract to closed chains of links---the first 
Betti number $b_1(\mathcal{T})$---counts its holes and so forth (see \cite{Hatcher,AlexandrovBook} and Section~\ref{section:methods}). 

\textbf{3. Dynamics of the coactivity complexes}. 
In practice, the coactivity complexes can be designed to reflect particular physiological properties of the cell assemblies. 
For example, the time course of the simplexes' appearance may reflect the dynamics of the cell assemblies' formation 
\cite{MWind1,MWind2,PLoZ,Hoffman}, or the details of the place cell activity modulations by the brain waves \cite{Arai,Basso}
and so on. In particular, a population of forming and disbanding cell assemblies can be represented by a set of appearing and 
disappearing simplexes, i.e., by a ``flickering'' coactivity complex $\mathcal{F}$ studied in\cite{MWind1,MWind2,PLoZ}. There 
it was demonstrated that if a cell assembly network rewires sufficiently slowly (tens of seconds to a minute timescale), then 
the ``topological shape'' of the corresponding coactivity complex remains stable and equivalent to the topology of the simulated
environment $\mathcal{E}$ shown on Fig.~\ref{Figure1}A, as defined by its Betti numbers $b_k(\mathcal{F})=b_k(\mathcal{E})=1$, 
$k = 0,1$ (Section~\ref{section:methods}). Physiologically, this implies that cell assemblies' turnover at the intermediate 
and the short memory timescales does not prevent the hippocampal network from producing a lasting representation of space, 
despite perpetual changes of its functional architecture \cite{Wang1}. 

\begin{figure}[!h]
	\includegraphics[scale=0.83]{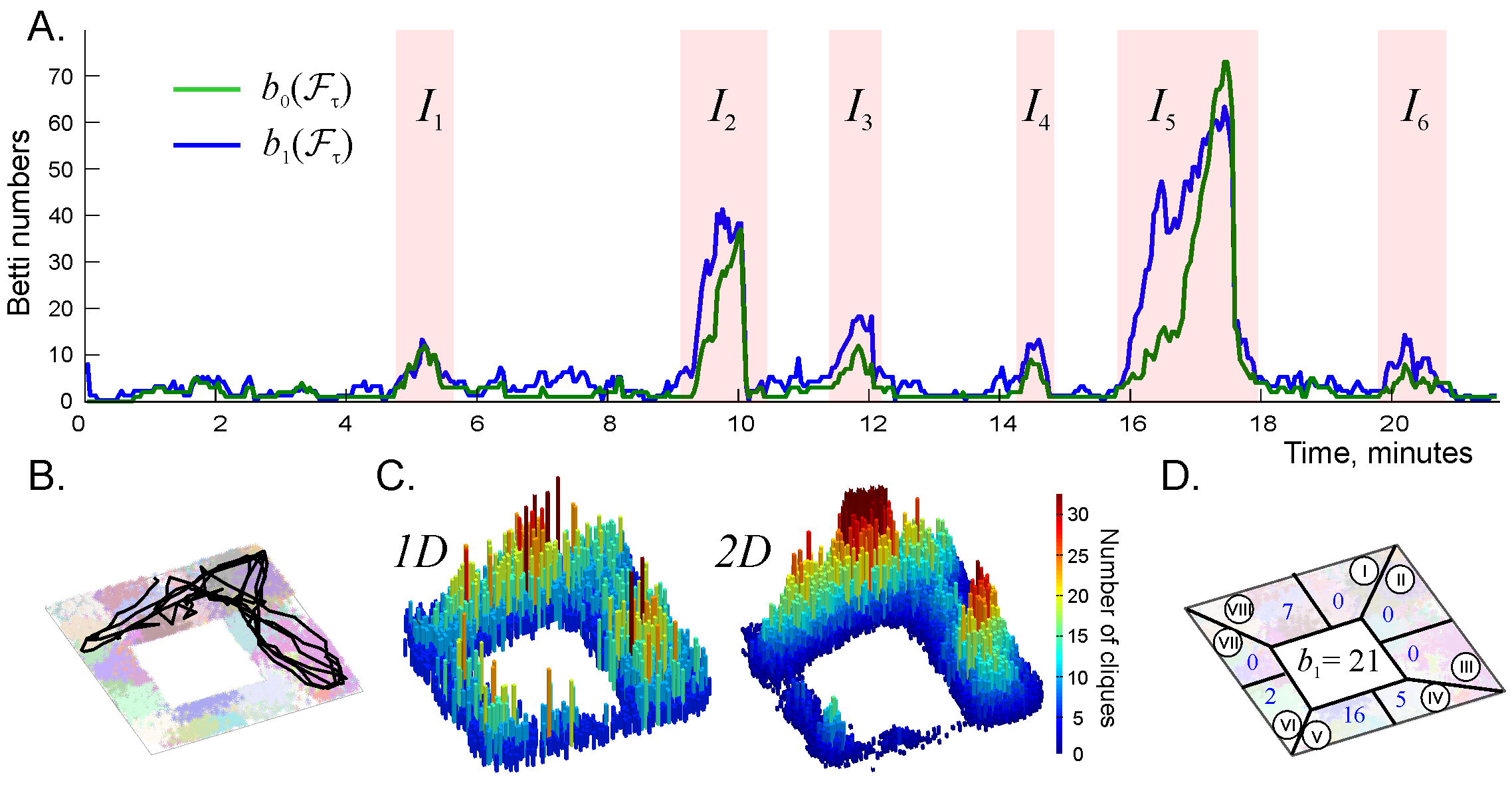}
	\caption{{\footnotesize\textbf{Topological fluctuations in a rapidly decaying coactivity complex in absence of replays.} 
			\textbf{A}. The green and the blue lines show, respectively, the zeroth and the first Betti numbers, $b_0(\mathcal{F}_{\tau})$ 
			and $b_1(\mathcal{F}_{\tau})$ (Section~\ref{section:methods}), as functions of time. For the most time, both Betti 
			numbers remain small, $\langle b_0(\mathcal{F}_{\tau})\rangle \approx 2.5$ and $\langle b_1(\mathcal{F}_{\tau}) \rangle
			\approx 2.8$, indicating a few disconnected fragments of the coactivity complex $\mathcal{F}_{\tau}$ and a few spurious 
			holes in them. The rapid increase of the Betti numbers during short ``instability intervals'' $I_1, \ldots, I_6$ 
			(highlighted by pink background) indicate periods of strong topological fluctuations in $\mathcal{F}_{\tau}$. 
			\textbf{B}. A segment of the simulated trajectory taken between the $16$th and $18$th minute shows that the animal spends 
			time before and during the instability period $I_5$ in a particular segment of the arena. During this time, the 
			connections over the unvisited segments of $\mathcal{E}$ start to decay (here the connections' mean lifetime is 
			$\tau = 50$ sec), as a result of which the coactivity complex $\mathcal{F}_{\tau}$ fractures into a large number 
			of disconnected pieces riddled in holes, which explains the splash of $b_0(\mathcal{F}_{\tau})$ and $b_1(\mathcal{F}_{\tau})$. 
			\textbf{C}. Spatial histograms of the links (i.e., centers of the pairwise place field overlaps, left panel) and of the 
			three-vertex simplexes (i.e., centers of triple place field overlaps, right panel) present in $\mathcal{F}_{\tau}$ 
			during the instability period $I_5$. The simplexes concentrate over the northeast corner of the environment, 
			whereas the populations of simplexes over the south and the southwestern parts thin out. 
			\textbf{D}. The ``local'' Betti number $b_1$ (blue numerals) computed separately for the eight sectors of the environment 
			(encircled Roman numerals) indicate that the holes emerge in all the ``abandoned'' parts, e.g., sector IV contains 
			5 holes and sector V contains 16 holes, etc. The global Betti number computed at about 16th minute for the entire 
			complex, $b_1(\mathcal{F}_{\tau}) = 21$, is shown in the middle.}}
	\label{Figure2}
\end{figure}

In particular, the model \cite{PLoZ} predicts that cell assembly network produces a stable topological map if the connections' 
mean lifetime exceeds $\tau \geq 150-200$ secs, which corresponds to the Hebbian plasticity timescale 
\cite{Billeh,Rakic,Hiratani,Russo,Zenke}. 
For noticeably shorter $\tau$, the topological fluctuations in the simulated hippocampal map are too strong and a 
stable representation of the environment fails to form. For example, in the case of the place field map shown on 
Fig.~\ref{Figure2}A, the connections' proper lifetime is about $\tau = 50$ secs and the corresponding coactivity 
complex is unstable: its Betti numbers frequently exceed the physical values ($b_k(\mathcal{F}) > b_k(\mathcal{E})$), 
implying that $\mathcal{F}$ may split into several disconnected pieces, each one of which may contain transient gaps, 
holes and other topological defects that do not correspond to the physical features of the environment. 

For the most time, these defects are scarce ($b_k(\mathcal{F}) < 5$) and may be viewed as topological irregularities that 
briefly disrupt otherwise functional cognitive map. Indeed, from the physiological perspective, it may be unreasonable to 
assume that biological cognitive maps never produce topological inconsistencies---in fact, admitting small fluctuations in 
a qualitatively correct representation of space may be biologically more effective than spending time and resources on 
acquiring a precise and static connectivity map, especially in dynamically changing environments. 
However, during certain periods, the topological fluctuations may become excessive, indicating the overall instability of 
the cognitive map. The origin of such occurrences is clear: if, e.g., the animal spends too much time in particular parts 
of the environment then the parts of $\mathcal{F}$ that represent the unvisited segments of space begin to deteriorate, 
leaving behind holes and disconnected fragments (Fig.~\ref{Figure2}B-D). 
Outside of these ``instability periods,'' when the rat regularly visits all segments of the environment, most place cells 
fire recurrently, thus preventing the coactivity complex $\mathcal{F}$ from deteriorating.

Although this description does not account for the full physiological complexity of synaptic and structural plasticity 
processes in the cell assembly network, it allows building a qualitative model that connects the animal's behavior, the 
parameters describing deterioration of the hippocampal network's functional architecture and the large-scale topological 
properties of the cognitive map. This, in turn, provides a context for testing the effects produced by the place cell 
replays, e.g., their alleged role in acquiring and stabilizing memories by strengthening the connections in parahippocampal 
networks \cite{Sadowski1,Sadowski2,Colgin}. To test these hypotheses, we adopted the topological model \cite{PLoZ} so 
that the decaying connections in the simulated hippocampal cell assemblies can be (re)established not only by the place 
cell activity during physical navigation but also by the endogenous activity of the hippocampal network, and studied the 
effect of the latter on the structure of the hippocampal map, as outlined below.

\textbf{4. The implementation of the coactivity complexes} is based on a classical model of the hippocampal network, in which 
place cells $c_i$ are represented as vertexes $v_i$ of a ``cognitive graph'' $\mathcal{G}$, while the connections between 
pairs of coactive cells are represented by the links, $\varsigma_{ij} = [v_i, v_j]$ of this graph \cite{Burgess,Muller,SchemaS}. 
The assemblies of place cells $\varsigma = [c_1, c_2, \ldots, c_n]$---the ``graphs of synaptically interconnected excitatory 
neurons,'' according to \cite{Syntax}---then correspond to fully interconnected subgraphs of $\mathcal{G}$, i.e., to 
its maximal cliques \cite{Hoffman,CAs,SchemaS}. Since each clique $\varsigma$, as a combinatorial object, 
can be viewed as a simplex spanned by the same set of vertexes (see Suppl. Fig. 6 in \cite{Basso}), the collection of cliques of 
the graph $\mathcal{G}$ defines a clique simplicial complex \cite{Jonsson}, which proves to be is one of the most successful
implementations of the coactivity complex. In previous studies \cite{Basso,Hoffman,CAs,SchemaS}, we demonstrated that in absence 
of decay ($\tau = \infty$), such a complex $\mathcal{T}$ effectively accumulates information about place cell coactivity at 
various timescales, capturing the correct topology of planar and voluminous environments. If the decay of the connections 
is taken into account ($\tau < \infty$), then the topology of the ``flickering'' coactivity complex $\mathcal{F}$ remains stable 
for sufficiently small rates, but if $\tau$ becomes too small, the topology of $\mathcal{F}$ may degrade. A question arises, 
whether the replays can slow down its deterioration, as the biological considerations suggest.

\textbf{5. Dynamics of the coactivity graph}. Physiologically, place cell spiking is synchronized with the components of the
extracellular local field potential---the so-called brain waves that also define the timescale of place cell coactivity \cite{BuzsakiBook}. 
Specifically, two or more place cells are considered coactive, if they fire spikes within two consecutive $\theta$-cycles---approximately 
150-250 msec interval \cite{Mizuseki}---a value that is also suggested by theoretical studies \cite{Arai}. In the following, this period 
will define the shortest timescale at which the functional connectivity of the simulated hippocampal network can change. For example,
a new link $\varsigma_{ij} = [v_i, v_j]$ in the coactivity graph will appear, if a coactivity of the cells $c_i$ and $c_j$ was detected 
during a particular $2\theta$ period. In absence of coactivity, the links can also disappear with probability
\begin{equation}
p_0(t) = \frac{1}{\tau}e^{-t/\tau},
\label{p0}
\end{equation}
where $t$ is the  time measured from the moment of last spiking of both cells $c_i$ and $c_j$ and the parameter $\tau$ defines 
the mean lifetime of the synaptic connections in the cell assembly network. In the following, $\tau$ will be the only parameter 
that describes the deterioration of the synaptic connections within the cell assemblies \cite{PLoZ}. We will therefore use the 
notations $\mathcal{G}_{\tau}$ and $\mathcal{F}_{\tau}$ to refer, respectively, to the flickering coactivity graph with decaying 
connections and to the resulting flickering coactivity complex with decaying simplexes. 

\textbf{6. Replays of place cell sequences} may in general represent both spatial and nonspatial memories. In the following, we 
simulate only spatial replays by constructing simplicial paths that represent previously navigated trajectories. Specifically, we 
select chains of connections that appeared in the coactivity graph $\mathcal{G}_{\tau}$ at the initial stages of navigation, and 
reactivate them at the later replay times $t_r$, $r = 1,2,\ldots,N_r$ \cite{Kudrimoti,ONeill1}. To replay a trajectory originating 
at a given timestep $t_i$, we randomly select a coactivity link $\varsigma_{kl}^{(i)} \in\mathcal{G}_{\tau}(t_i)$ that is active 
within that time window; this link is then gives rise to a sequence of joined links, randomly selected among the ones that activate 
at the consecutive time steps, $\varsigma_{lm}^{(i+1)}, \varsigma_{mn}^{(i+2)}, \ldots$. Since there are typically several active 
links at every moment, this procedure allows generating a large number of replay trajectories. The physiological duration of 
replays---typically about 100-200 msec \cite{Colgin}---roughly corresponds to the coactivity window widths, i.e., to the timesteps 
in which the coactivity graph evolves; we therefore ``inject'' the activated links into a particular coactivity window $t_r$ in 
order to simulate rapid replays.

After a simplicial trajectory is replayed, the injected links begin to decay and to (re)activate in the course of the animal's moves 
across the environment, just as the rest of the links. Most of these ``reactivated'' links simply rejuvenate the existing connections
in $\mathcal{G}_{\tau}$. However, some injections instantaneously reinstate decayed connections and produce an additional population 
of higher order cliques, which affect the topological properties of the coactivity complex $\mathcal{F}_{\tau}$, and hence---according 
to the model---of the cognitive map. As mentioned previously, hippocampal replays are believed to enable spatial learning by stimulating 
inactive connections, by slowing down their decay and by reinforcing cell assemblies' stability \cite{Carr,Jadhav1,Sadowski2,Ego,
	Girardeau1,Girardeau2}. 
In the model's terms, this hypothesis translates as follows: the additional influx of rejuvenated simplexes provided by the replays 
should qualitatively improve the topological structure of the flickering coactivity complex, slow down deterioration of its simplexes, 
suppress its topological defects and in general help to sustain its topological integrity. In the following, we test this hypothesis by 
simulating different patterns of the place cell reactivations and quantifying the effect that this produces on the simulated cognitive map.

\section{Results}
\label{section:results}

\begin{figure}[!h]
	\includegraphics[scale=0.83]{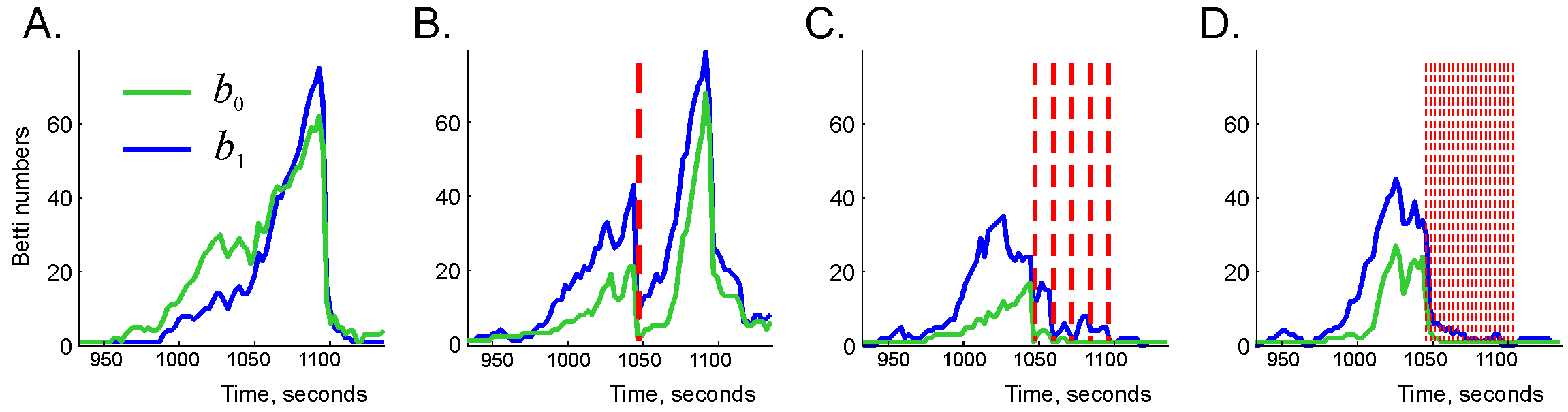}
	\caption{{\footnotesize\textbf{Suppressing topological fluctuations by reactivation of simplexes.} 
			\textbf{A}. During the instability period $I_5$ (approximately between 15.5 and 18.5 mins) the topological fluctuations 
			in the coactivity complex become very strong, with the Betti numbers soaring at $b_0(\mathcal{F}_{\tau})\approx 65$ and 
			$b_1(\mathcal{F}_{\tau})\approx 75$. 
			\textbf{B}. If all the reactivated links are injected into the coactivity complex at once, at a moment preceding the peak 
			of the Betti numbers (marked by a vertical red dashed line), the fluctuations in the coactivity complex are immediately 
			suppressed. However, as the connection decay takes over, the fluctuations kick back, reaching the original high values in 
			under a minute. 
			\textbf{C}. Five consecutive replays, marked by five vertical red dashed lines, produce a more lasting effect, reducing the 
			Betti numbers to smaller values $\langle b_0(\mathcal{F}_{\tau})\rangle \approx 3$ and $\langle b_1(\mathcal{F}_{\tau})
			\rangle \approx 7$ over the remainder of the instability period. 
			\textbf{D}. More frequent replays (once every $2.5$ secs, vertical dashed lines) nearly suppress the topological fluctuations, 
			producing the average values $\langle b_0(\mathcal{F}_{\tau})\rangle\approx 1.2$ and $\langle b_1(\mathcal{F}_{\tau})\rangle
			\approx 3$, i.e., 
			leaving only a couple of spurious loops in $\mathcal{F}_{\tau}$.}}
	\label{Figure3}
\end{figure}

\textbf{1. Initial testing}. The effect produced by the replays on the cognitive map depends on the parameters of the model: the 
selection of replayed trajectories, the injection times, the frequency of the replays and so forth. To start the simulations, we 
selected $N_s = 80$ different replay sequences originating at $N_i = 25$ moments of time, $t_i$, $i = 1,\ldots, N_i$, between $20$ 
and $200$ seconds of navigation (the initial interval $I_{init}$). During this period, the trajectory covers the arena more or less 
uniformly: a typical $25$ secs long segment of a trajectory extends across the entire the environment and contains on average about 
$l_s = 100$ links (Fig.~\ref{SFigure1}). As a result, the corresponding simplicial paths traverse the full coactivity complex 
$\mathcal{F}_{\tau}$ and one would expect that replaying these paths should help to suppress the topological defects in $\mathcal{F}$. 
To verify this prediction, we replayed the resulting pool of $\mathcal{G}$-link sequences within the main instability period $I_5$ 
using different approaches and tested whether this can suppress the topological fluctuation (Fig.~\ref{Figure3}A).

In the first scenario, all replay chains were injected into the connectivity graph $\mathcal{G}_{\tau}$ at once, in the middle of the 
instability period $I_5$ (Fig.~\ref{Figure3}B). As a result of such a ``massive'' instantaneous replay, the topological fluctuations are 
initially suppressed but then they quickly rebound, producing about the same number of spurious $0D$ loops (i.e., the cognitive map 
remains as fragmented as before) and an even higher number of $1D$ loops that mark spurious holes in the cognitive map (Suppl. 
Movie 1). In other words, our model suggests that a single ``memory flash'' fails to correct the deteriorating memory
map even at a short timescale, which suggests that more regular replay patterns are required.

Indeed, if the same set of replay sequences is uniformly distributed into $N_r = 5$ consecutive groups inside the instability period 
$I_5$ (one group per 36 seconds, $N_s/N_r = 16$ chains of links each), then the topological fluctuations in the coactivity complex
$\mathcal{F}_{\tau}$ subside more and over a longer period (see Fig.~\ref{Figure3}C and Suppl. Movie 2). If the 
replays are produced even more frequently (every 9 seconds, i.e., about 20 replays total, $N_s/N_r \approx 4$ chains of links injected 
per replay) then the topological fluctuations in $I_5$ are essentially fully suppressed over the entire environment (Fig.~\ref{Figure3}D, 
Fig.~\ref{SFigure2} and Suppl. Movie 3).

One can draw two principal observations from these results: first, that spontaneous reactivation of connections at the physiological 
timescale can qualitatively alter the topological structure of the flickering coactivity complex and second, that the temporal pattern 
of replays plays a key role in suppressing the topological fluctuations in the cognitive map. 

\textbf{2. Implementation of the replays}. Electrophysiological data shows that the frequency of the replays ranges between 
$0.1$ Hz in active navigation to $0.4$ Hz in quiescent states and $4$ Hz during sleep \cite{ONeill1,Colgin,Jadhav1,Sadowski2}.
Since we model spatial learning taking place during active navigation, we implemented replays at the maximal rate of $0.4$ Hz, 
which corresponds to no more than one replay event over ten consecutive coactivity intervals. Second, we took into account the 
fact that, in complex environments, hippocampus may replay a few sequences simultaneously. For example, on the $Y$-track 
\cite{ONeill1} two simultaneous replay sequences can represent the two prongs of the $Y$. In open environments, there may 
be more simultaneously replayed sequences; however we used the most conservative estimate and replayed two different 
sequences at each replay moment $t_r$.

\begin{figure}[!h]
	\includegraphics[scale=0.83]{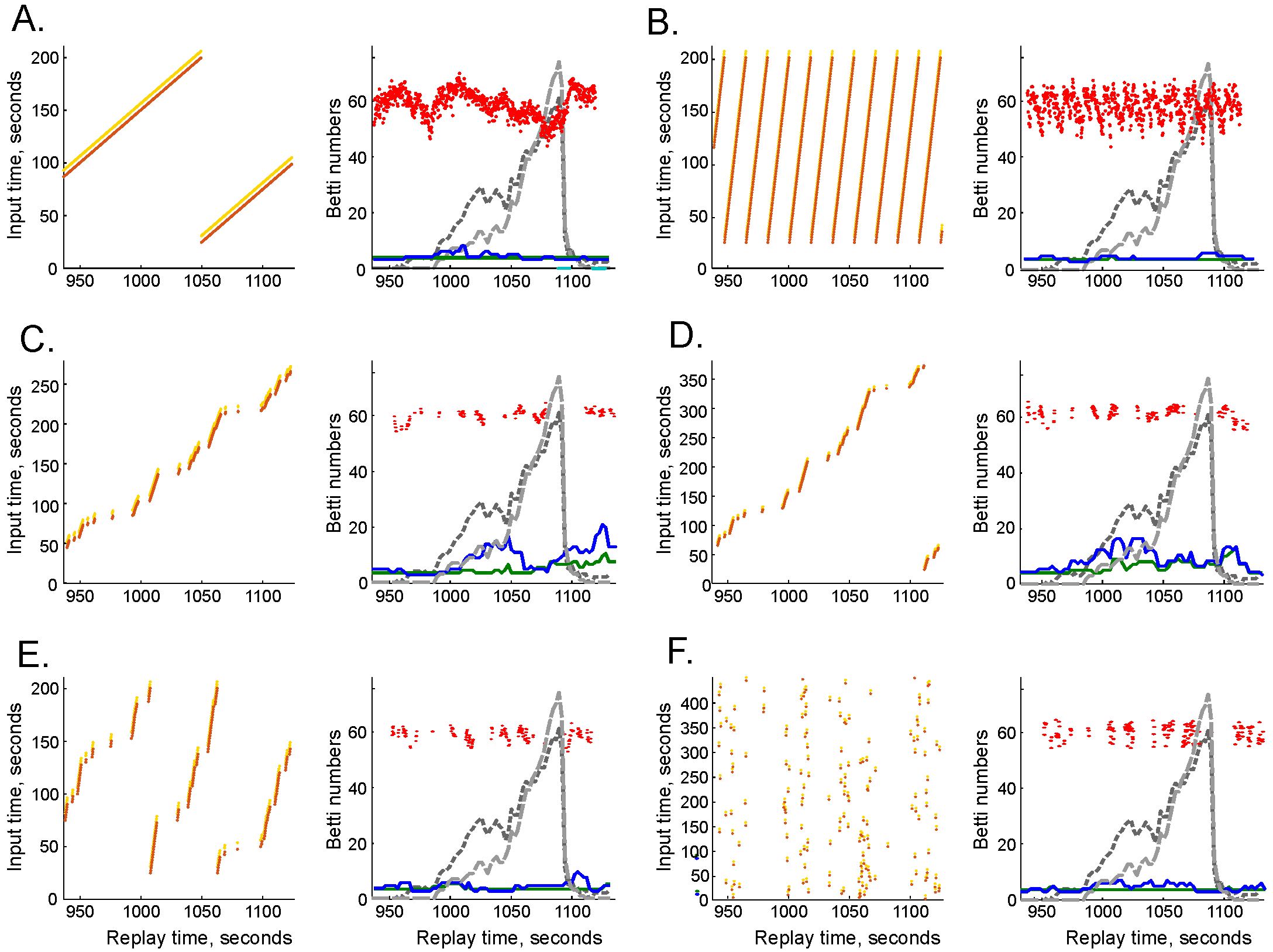}
	\caption{{\footnotesize\textbf{Suppressing the topological fluctuations by replays.} In all cases, the injection diagram 
			on the left relates the times when the place cell sequences were originally produced (vertical axis) to the times 
			when they are replayed (horizontal axis). Each yellow and brown dot corresponds to a replayed sequence. 
			On the right panel, the replay times are marked by red dots, with the vertical scatter proportional to the simulated 
			speed of the animal. The resulting zeroth Betti number ($b_0(\mathcal{F}_{\tau})$, green line) and the first Betti 
			number ($b_1(\mathcal{F}_{\tau})$, blue line) are shown in the foreground, and the original, unstable Betti numbers 
			(without replays) are shown in the background (dark and light dashed gray lines, respectively). 
			\textbf{A}. A simple translational replay of a couple of sequences repeated over a three minute period (between $20$ 
			and $200$ secs), with a $\Delta T = 14$ min delay. The zeroth Betti number regains its physical value, $b_0(\mathcal{F}_{\tau}) 
			= b_0(\mathcal{E}) = 1$, indicating that cognitive map reconnects into one piece. The first Betti number fluctuates 
			near the physical value $b_1(\mathcal{F}_{\tau}) = 1.5 \pm 2.2$, indicating that nearly all spurious holes are closed. 
			\textbf{B}. Compressed replay: a three-minute period is replayed repeatedly over several consecutive $20$ sec intervals. Here 
			the zeroth Betti number remains correct, $b_0(\mathcal{F}_{\tau}) = 1$, and the fluctuations of the first Betti number 
			reduce farther, $b_1(\mathcal{F}_{\tau}) = 1.2 \pm 1$. 
			C. Modulating the replay times by slow move periods ($v < 15$ cm/sec) produces sparser replays. As a result, the 
			topological fluctuations in the case of uncompressed, speed-modulated delayed replays increase, $b_0(\mathcal{F}_{\tau}) 
			= 4.2 \pm 1.9$, $b_1(\mathcal{F}_{\tau}) = 8.2 \pm 3.4$.
			\textbf{D}. Small compressions (up to $300$ seconds replayed over $\sim 150$ second period, same delay) may intensify the 
			fluctuations: $b_0(\mathcal{F}_{\tau}) = 5.9 \pm 2.4$, $b_1(\mathcal{F}_{\tau}) = 9.2 \pm 4.1$. 
			\textbf{E}. Further compression of the replays improves the results, $b_0(\mathcal{F}_{\tau}) = 1.1 \pm 1.4$, $b_1(\mathcal{F}_{\tau}) 
			= 1.7 \pm 2$, although the variations of $b_1(\mathcal{F}_{\tau})$ remain high compared to the cases in which the 
			replays are not modulated by the speed. 
			\textbf{F}. Random replays reduce the fluctuations even further: the coactivity complex acquires the correct zeroth Betti number 
			$b_0(\mathcal{F}_{\tau}) = 1$ (the map becomes connected), producing occasional spurious loops, $b_1 = 1.4 \pm 
			1.2$---occasional topological irregularities. 
			}}
	\label{Figure4}
\end{figure}

In the simplest scenario, we injected pairs of sequences into the coactivity graph $\mathcal{G}_{\tau}$ with a constant delay of 
about $\Delta T = t_i - t_r\approx 14$ minutes after their physical onset, which placed them inside of the instability period $I_5$ 
(Fig.~\ref{Figure4}A). In response, the topological fluctuations in the coactivity complex $\mathcal{F}_{\tau}$ significantly 
diminished. In fact, the zeroth Betti number (the number of the disconnected components) regained its physical value 
$b_0(\mathcal{F}_{\tau}) = 1$, indicating that the replays helped to pull the fragments of the cognitive map together into a single 
connected piece. The first Betti number (the number of holes) remains on average close to its physical value, $\langle 
b_1(\mathcal{F}_{\tau})\rangle  = 1.5$, exhibiting occasional fluctuations, $\Delta b_1=\pm 2.2$.

As mentioned above, the occasional islets separating from the main body of the simplicial complex or a few small holes appearing 
in it for a short period should be viewed as topological irregularities, rather than signs of topological instability. We therefore base 
the following discussion on addressing only the qualitative differences produced by the replays on the topology of the cognitive map: 
whether replays can prevent fracturing of the complex into multiple pieces and rapid proliferation of spurious loops in all dimensions.
From such perspective, our results demonstrate that translational replays at a physiological rate can effectively restore the correct
topological shape of the cognitive map, which illustrates functional importance of the replay activity.

Since the replays are generated by the endogenous activity of the hippocampal network, the relative temporal order of the replayed 
sequences can be altered, i.e., the replay times $t_r$ can be spread wider or denser than their ``physical'' origination times $t_i$. 
The effect of the replays will be, respectively, weaker or stronger than in the case of translational delay, due to the corresponding 
changes of the sheer number of the reactivated links. However, one can factor out the direct contribution of the replays' volume and 
study more subtle effects produced specifically by the replay's temporal organization. 
To this end, we split the replay period $I_5$ into a set of $N_R$ shorter subintervals, $I^{1}_5, I^{2}_5,\ldots I^{N_R}_5$, and then 
replayed the sequences of links originating from the initial three minute interval $I_{init}$ within each subinterval $I^{n}_5$, 
$n=1,2,\ldots, N_R$. Since only two sequences are replayed within every coactivity window, the total number of the (re)activated 
sequences remains the same as in the delayed replay case, even though the source interval $I_{init}$ is 
compressed $N_R$-fold in time. Thus, the difference between the effects produced by the ``compressed'' replays will be due solely 
to the differences in their temporal reorganizations. 

The results illustrated on Fig.~\ref{Figure4}B demonstrate that the compressed replays suppress the topological fluctuations more 
effectively. For example, the repeated replay in a sequence of 20 sec intervals ($N_R = 9$ fold compression) not only restores the 
correct value of the zeroth Betti number, $b_0(\mathcal{F}_{\tau}) = 1$, but also drives the average number of noncontractible 
simplicial loops close to physical value, $\langle b_1(\mathcal{F}_{\tau})\rangle  = 1.2 \pm 1$. In other words, $\mathcal{F}_{\tau}$ 
almost regains its topologically correct shape, with an occasional spurious hole appearing for less than a second.
Physiologically, these results suggest that time-compressed, repetitive ``perusing'' through memory sequences helps to prevent  
deterioration of global memory frameworks better than simple ``orderly'' recalls. 

\textbf{3. Speed modulation of the replays}. Since replays are mostly observed during quiescent periods and slow moves 
\cite{ONeill1,ONeill2}, we studied whether such ``low-speed'' replays will suffice for suppressing the topological fluctuations in the 
cognitive map. Specifically, we identified the periods when the speed of the animal falls below $15$ cm/sec (which, in our simulations 
happens during $14\%$ of time, see Fig.~\ref{SFigure3} and \cite{Nadasdy,Koene}), and replayed the place cell sequences only 
during these periods.

It turns out that although the resulting slow motion replays can stabilize the topological structure of the simulated cognitive map, 
the effect strongly depends on their temporal organization. Specifically, in the simple delayed replay scenario, the topological 
fluctuations remain significantly higher than without speed modulation (Fig.~\ref{Figure4}C, Suppl. Movie 4). On average, the coactivity 
complex contains about a dozen of spurious loops: it remains split in a few pieces, $\langle b_0(\mathcal{F}_{\tau})\rangle = 4.2$, that
together contain $\langle b_1(\mathcal{F}_{\tau})\rangle = 8.2$ holes on average. This is a natural result---one would expect that
speed restrictions will diminish the number of the injected active connections and hence that $\mathcal{F}_{\tau}$ will degrade more. 
A slightly compressed replay (4 minutes of activity replayed over 3 minute period) does not improve the result: both the number 
of disconnected components and the number of holes in them increase $\langle b_0(\mathcal{F}_{\tau})\rangle  = 5.9$, $\langle
b_1(\mathcal{F}_{\tau})\rangle = 9.2$ (Fig.~\ref{Figure4}D). However, if the replays are compressed further, the average number
of disconnected components is significantly reduced: for the threefold compression shown on Fig.~\ref{Figure4}E, the mean values 
are $\langle b_0(\mathcal{F}_{\tau})\rangle  = 1.1$ and $\langle b_1(\mathcal{F}_{\tau})\rangle  = 1.7$, i.e., the encoded map 
approaches the quality of the maps produced with unrestricted replays. 

\begin{figure}[!h]
	\includegraphics[scale=0.83]{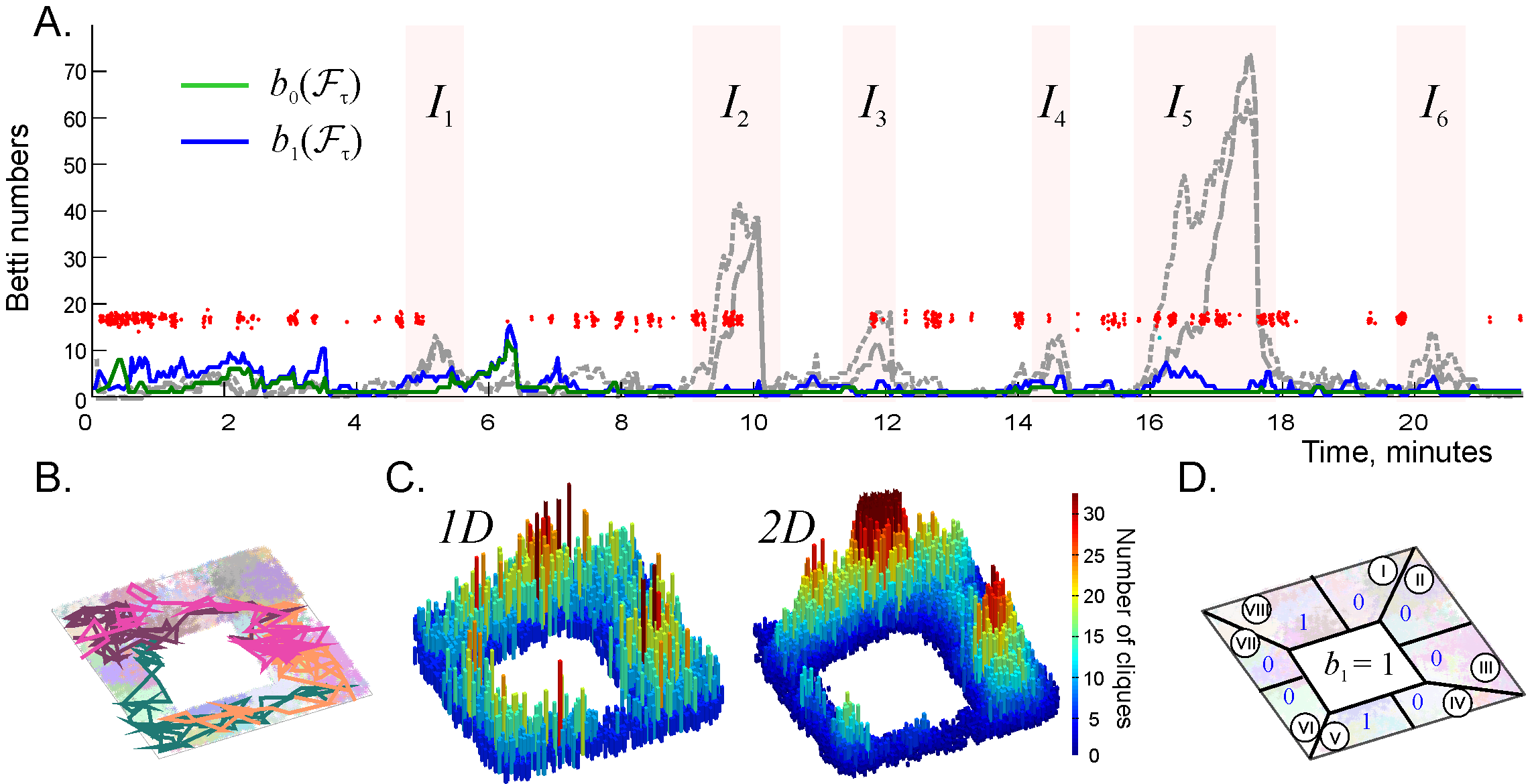} 
	\caption{{\footnotesize\textbf{Replays suppress topological fluctuations over the entire navigation period.} 
			\textbf{A}. The time dependence of the Betti numbers $b_0(\mathcal{F}_{\tau})$ and $b_1(\mathcal{F}_{\tau})$ (the green 
			and the blue line respectively) in presence of the replays. The topological fluctuations that previously overwhelmed the 
			map during the ``instability intervals'' (shown in the background by two dashed gray lines, see Fig.~\ref{Figure2}A) 
			are now nearly fully suppressed. The replay moments, $t_r$, marked by the red dots, are modulated by speed, $v < 15$ 
			cm/sec.
			\textbf{B}. Several examples of replayed trajectories over the navigated environment are shown in different colors (see also 
			Fig.~\ref{SFigure1}). 
			\textbf{C}. Spatial histograms of the centers of the links (left panel) and of the three-vertex simplexes (right panel) during 
			the instability period $I_5$ in presence of the replays. The populations of simplexes over the south and the southwestern 
			parts of the environment in presence of the replays have increased compared to the case shown on Fig.~\ref{Figure2}C, 
			which suppresses spurious topological loops in the coactivity complex. 
			\textbf{D}. The Betti numbers $b_1$ computed for the eight sectors of the environment are also significantly reduced, indicating 
			that the topological fluctuations are suppressed both locally and globally. The deviations of $b_1$ from 0 in the sectors 
			V and VIII are due to boundary effects at the sector's edges that do not affect the global value $b_1(\mathcal{F}_{\tau}) 
			= 1$. All zeroth Betti numbers, both local and global, assume correct values $b_0 = 1$ and are not shown.}}
	\label{Figure5}
\end{figure}

The effectiveness of the latter scenario can be explained by noticing that replay compression brings the activities that are widely 
spread in physical time into close temporal vicinities during the replays. In other words, in compressed replays, a wider variety 
of connections is activated at each $t_r$: the real-time separation between activity patterns shrinks. This helps to reduce or eliminate 
the temporal ``lacunas'' in place cell coactivity across the entire hippocampal network and hence to prevent spontaneous deterioration 
of its parts. In physiological terms, this implies that the compressed replays of the place cell patterns are less constrained by the 
physical temporal scale of the rat's navigational experiences, which leads to a more even activation of the connections in the network 
and helps to prevent the memory map's fragmentation.

To test this idea, we amplified this effect by shuffling the order of the replayed sequences and by randomizing the injection 
diagram, thus enforcing a nearly uniform pattern of injected activity across the simulated population of cell assemblies. This indeed
proved to be the most effective replay strategy: as shown on Fig.~\ref{Figure4}F, such replay patterns restore the topological shape 
of the coactivity complex, allowing only occasional holes: $\langle b_0(\mathcal{F}_{\tau})\rangle=1$, $\langle b_1(\mathcal{F}_{\tau})
\rangle\approx 1.4\pm 1.2$ (Suppl. Movie 5). Thus, the model suggests that ``reshuffling'' the temporal sequence of memory replays helps 
to sustain memory framework better than orderly recollections, occurring in natural past-to-future succession.
The effect of random replays of the place cell sequences over the entire simulated navigation period shown on Fig.~\ref{Figure5} 
clearly illustrates the importance of replays for rapid encoding of topological maps: the fluctuations in the cognitive map are uniformly 
suppressed.

\section{Discussion}
\label{section:discuss}

The model discussed above suggests that replays of place cell activity help to learn and to sustain the topological structure of the 
cognitive map. The physiological accuracy of the replay simulation can be increased \textit{ad infinitum}, by incorporating more and 
more parameters into the model. In this study we use only a few basic properties of the replays, which, however, capture several key 
functional aspects of the replay activity. First, the model implements an effective feedback loop, in which the onset of topological 
instabilities in the flickering coactivity complex $\mathcal{F}_{\tau}$ triggers  the replays that restore its integrity. Indeed, the 
cell assemblies' (and the corresponding simplexes') decays intensify as the animal's exploratory movements slow down and visits to 
particular segments of the environment become less frequent. On the other hand, low speed periods define temporal windows during which 
the simulated replays are injected into the network, which work to suppress the topological instabilities.
Second, the model allows controlling the replays' temporal organization independently from the other parameters or neuronal activity 
and exploring the replays' contribution into acquiring and stabilizing the cognitive maps. The results demonstrate that in order to 
strengthen the decaying connections in the hippocampal network effectively, the replays must 1) be produced at a sufficiently high rate
that falls within the physiological range and 2) distribute without temporal clustering, in a semi-random order.

An important aspect of the obtained results is a separation of the timescales at which different types of topological information is 
processed. On the one hand, rapid turnover of the information about local connectivity at the working memory timescale is represented 
by quick recycling of the cell assemblies and rapid spontaneous replays of the learned sequences. On the other hand, the large-scale 
topological structures of the cognitive map, described by the instantaneous homological characteristics of the coactivity complex, 
emerge at the intermediate memory timescale. Thus, the model suggests that the characteristic timescale of the topological loops' dynamics 
is by an order of magnitude larger than the timescale of fluctuations at the cell assembly level. This observation provides a functional 
perspective on the role played by the place cell replays in learning: by reducing the fluctuations, replays help separating the fast and 
the slow information processing timescales and hence to extract stable topological information that can be used to build a long-term, 
qualitative representation of the environment. This separation of timescales corroborates with the well-known observation that transient 
information is rapidly processed in the hippocampus and then the resulting memories are consolidated and stored in the cortical areas, 
but at slower timescales and for longer periods. 

\section{Methods}
\label{section:methods}

\paragraph{Topological Glossary.} For the reader's convenience, we briefly outline the key topological terms and concepts used 
in this paper.

$\bullet$
\textit{An abstract simplex} of order $d$ is a set of $(d +1)$ elements, e.g., a set of coactive cells, $\sigma^{(d)}=[c_{i_0},
c_{i_1}, \ldots, c_{i_d}]$ or a set of place fields, $\sigma^{(d)}=[\upsilon_{j_0},\upsilon_{j_1},\ldots,\upsilon_{j_k}]$. The 
subsets of $\sigma^{(d)}$ are its \textit{subsimplexes}. Subsimplexes of maximal dimensionality $(d-1)$ are referred to as 
\textit{facets} of $\sigma^{(d)}$. 

$\bullet$
\textit{An abstract simplicial complex} $\Sigma$ is a family of abstract simplexes closed under the overlap relation: a nonempty 
overlap of any two simplexes $\sigma^{(d_1)}_1\in\Sigma$ and $\sigma^{(d_2)}_2\in\Sigma$ is a subsimplex of both $\sigma^{(d_1)}_1$ 
and $\sigma^{(d_2)}_2$. 

$\bullet$
Geometrically, simplexes can be visualized as $d$-dimensional polytopes: $\sigma^{(0)}$ as a point, $\sigma^{(1)}$ as a line segment, 
$\sigma^{(2)}$ as a triangle, $\sigma^{(3)}$ as a tetrahedron, etc.  
The corresponding \textit{geometric simplicial complexes} are multidimensional polyhedra that have a shape and a \textit{structure} 
that does not change with simplex deformations, e.g., disconnected components, holes, cavities of different dimensionality, etc. This 
structure, commonly referred to as \textit{topological} \cite{AlexandrovBook}, is identical in a geometric simplicial complex to and 
in the abstract complex built over the vertexes of the geometric simplexes. Thus, abstract simplicial complexes may be viewed as
structural \textit{representations} of the conventional geometric shapes.

$\bullet$
Topological properties of the simplicial complexes are established based on algebraic analyses of chains, cycles and boundaries. 

\textit{A chain} $\alpha^{(d)}$ is a formal combination $d$-dimensional simplexes with coefficients from an algebraic ring or a field. 
Intuitively, they can be viewed, e.g., as the simplicial paths described in Section~\ref{section:model}. Such combinations permit 
algebraic operations: they can be added, subtracted, multiplied by a common factor, etc. As a result, the set of all chains of a given 
simplicial complex, $C(\Sigma)$, also forms an algebraic entity, e.g., if the chains' coefficients form to a field, then $C(\Sigma)$ 
forms a vector space.

\textit{A boundary} of a chain, $\partial\alpha^{(d)}$, is a formal combination of all the facets of the $\alpha$-chain, with the 
coefficients inherited from $\alpha$ and alternated so that the boundary of $\partial\alpha^{(d)}$ vanishes, $\partial^2\alpha^{(d)}=0$. 
This universal topological principle---boundary of a boundary is a null set---can be illustrated on countless examples, e.g., by noticing 
that the external surface of a triangular pyramid $\sigma^{(3)}$---its geometric boundary---has no boundary itself.  

\textit{Cycles} generalize the previous example---a generic cycle $z$ is a chain without a boundary, $\partial z=0$. Intuitively, cycles 
correspond to agglomerates of simplexes (e.g., simplicial paths) that loop around holes and cavities of the corresponding dimension. Note 
however, that although all boundaries are cycles, not all cycles are boundaries.

$\bullet$
\textit{Homologies}. Two cycles, $z_1$ and $z_2$, are \textit{equivalent}, or \textit{homologous}, if they differ by a boundary chain. 
The set of equivalent cycles forms a \textit{homology class}. If the chain coefficients come from a field, then the homology classes of 
$d$-dimensional cycles form a vector space $\Hgr_d(\Sigma)$. The dimensionality of this vector space is the $d$-th \textit{Betti number} 
of the simplicial complex $\Sigma$, $b_d(\Sigma) = \dim \Hgr_k(\Sigma)$, which counts the number of independent $d$-dimensional holes 
in $\Sigma$.

$\bullet$
\textit{Flickering complexes} $\mathcal{F}(t)$ consist of simplexes that may disappear or (re)appear, so that the complex as a whole may 
grow or shrink from one moment to another (see Fig.2 in \cite{PLoZ}),
$$\mathcal{F}(t_1)\subseteq\mathcal{F}(t_2)\subseteq\mathcal{F}(t_3)\supseteq\mathcal{F}(t_4)\subseteq\mathcal{F}(t_5)\supseteq \ldots.$$
Computing the corresponding Betti numbers, $b_k(\mathcal{F}(t))$, requires a special technique---\textit{Zigzag persistent homology} theory 
that allows tracking cycles in $\mathcal{F}$ on moment-to-moment basis \cite{Carlsson2,Carlsson3,PLoZ}.

$\bullet$
\textit{A clique} in a graph $G$ is a set of fully interconnected vertices, i.e., a complete subgraph of $G$. Combinatorially, cliques 
have the same key property as the abstract simplexes: any subcollection of vertices in a clique is fully interconnected. Hence a nonempty 
overlap of two cliques $\varsigma$ and $\varsigma'$ is a subclique in both $\varsigma$ and $\varsigma'$, which implies that cliques may 
be formally viewed as abstract simplexes and a collection of cliques in a given graph $G$ produces its \textit{clique simplicial complex} 
$\Sigma(G)$ \cite{Jonsson}. In particular, the \textit{clique coactivity complexes} $\mathcal{T}_{\varsigma}$ is induced from the 
\textit{coactivity graphs} $\mathcal{G}$ \cite{Basso,Hoffman,CAs} and the flickering clique complexes $\mathcal{F}_{\tau}$ are constructed 
using coactivity graph with flickering connections $\mathcal{G}_{\tau}$, \cite{MWind1,MWind2,PLoZ}. Note however, that the topological 
analyses address the topology of the coactivity complexes, rather than the network topology of $\mathcal{G}$.

\paragraph{Spike simulations.}
The environment  shown on Fig.~\ref{Figure1}A is simulated after typical arenas used in typical electrophysiological experiments. Over 
the navigation period $T_{tot} = 30$ min, the trajectory covers the environment uniformly. The maximal speed of the simulated movements 
is $v_{\max} = 50$ cm/sec, with the mean value $\bar v = 25$ cm/sec. The firing rate of a place cell $c$ is defined by
\begin{equation}
\lambda_c(r)=f_c e^{-\frac{(r-r_c)^2}{2s^2_c}}
\nonumber
\end{equation}
where $f_c$ is the maximal firing rate and $s_c$ defines the size of the place field centered at $r_c$ \cite{Barbieri}. In addition, 
spiking is modulated by the $\theta$-oscillations----a basic cycle of the extracellular local field potential in the hippocampus, with the 
frequency of about $8$ Hz \cite{Arai,Mizuseki,Huxter}. The simulated ensemble contains $N_{c} = 300$ virtual place cells, with the typical 
maximal firing rate $f = 14$ Hz and the typical place field size $s = 20$ cm. 

\section{Acknowledgments}
\label{section:acknow}

The work was supported by the NSF 1422438 grant.

\newpage
\beginsupplement

\section*{SUPPLEMENTARY FIGURES}

\begin{figure}[ht]
	\includegraphics[scale=0.83]{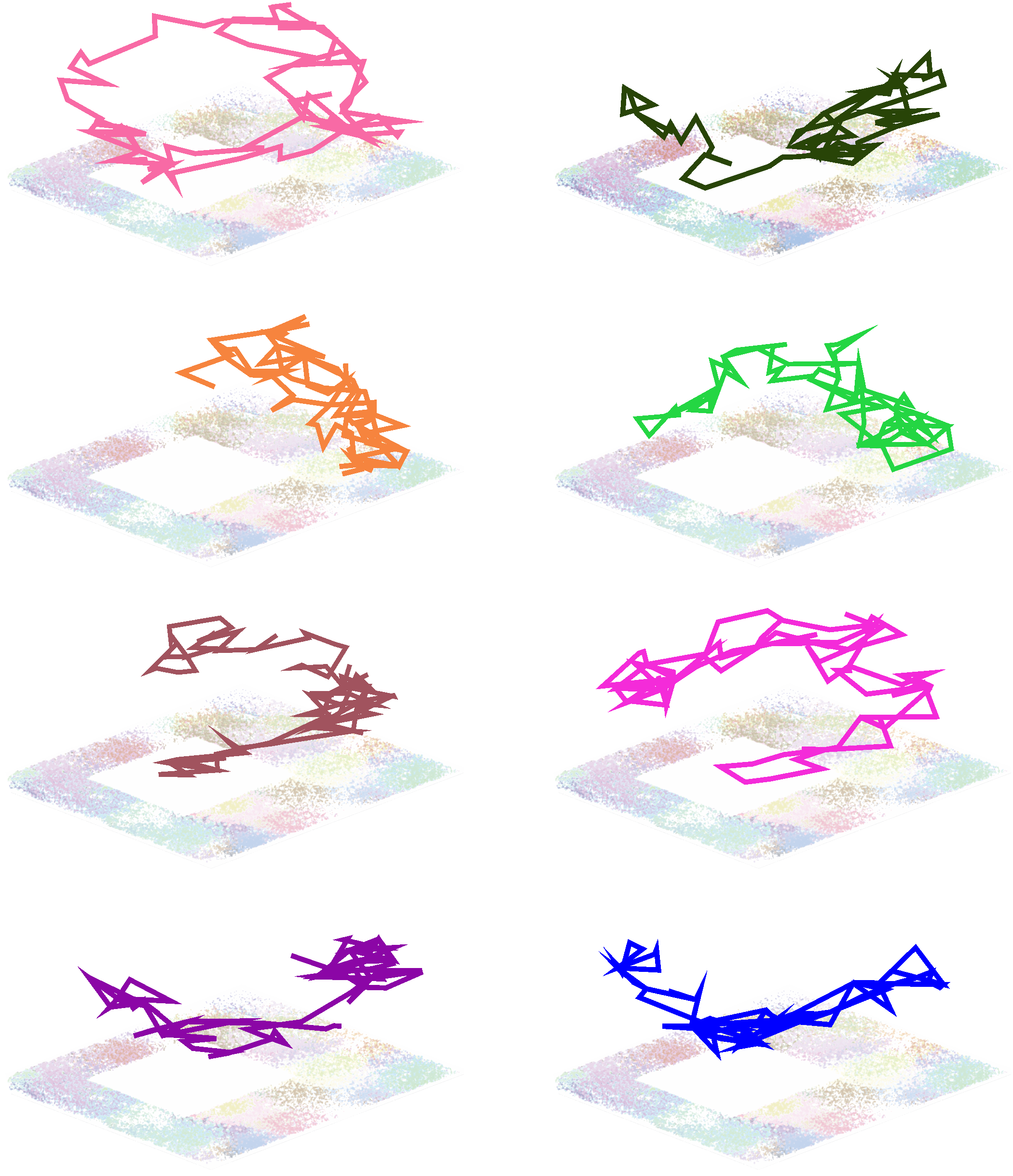}
	\caption{{\footnotesize \textbf{Replayed trajectories.} Eight simulated trajectories of length $l_s = 100$, which corresponds to about
			$25$ secs of of physical time, extending across different segments of the environment. The positions of the replayed trajectories 
			over the environment are not related to the location of the rat at the moment when the replays occurred.}}
	\label{SFigure1}
\end{figure} 


\begin{figure}[ht]
	\includegraphics[scale=0.8]{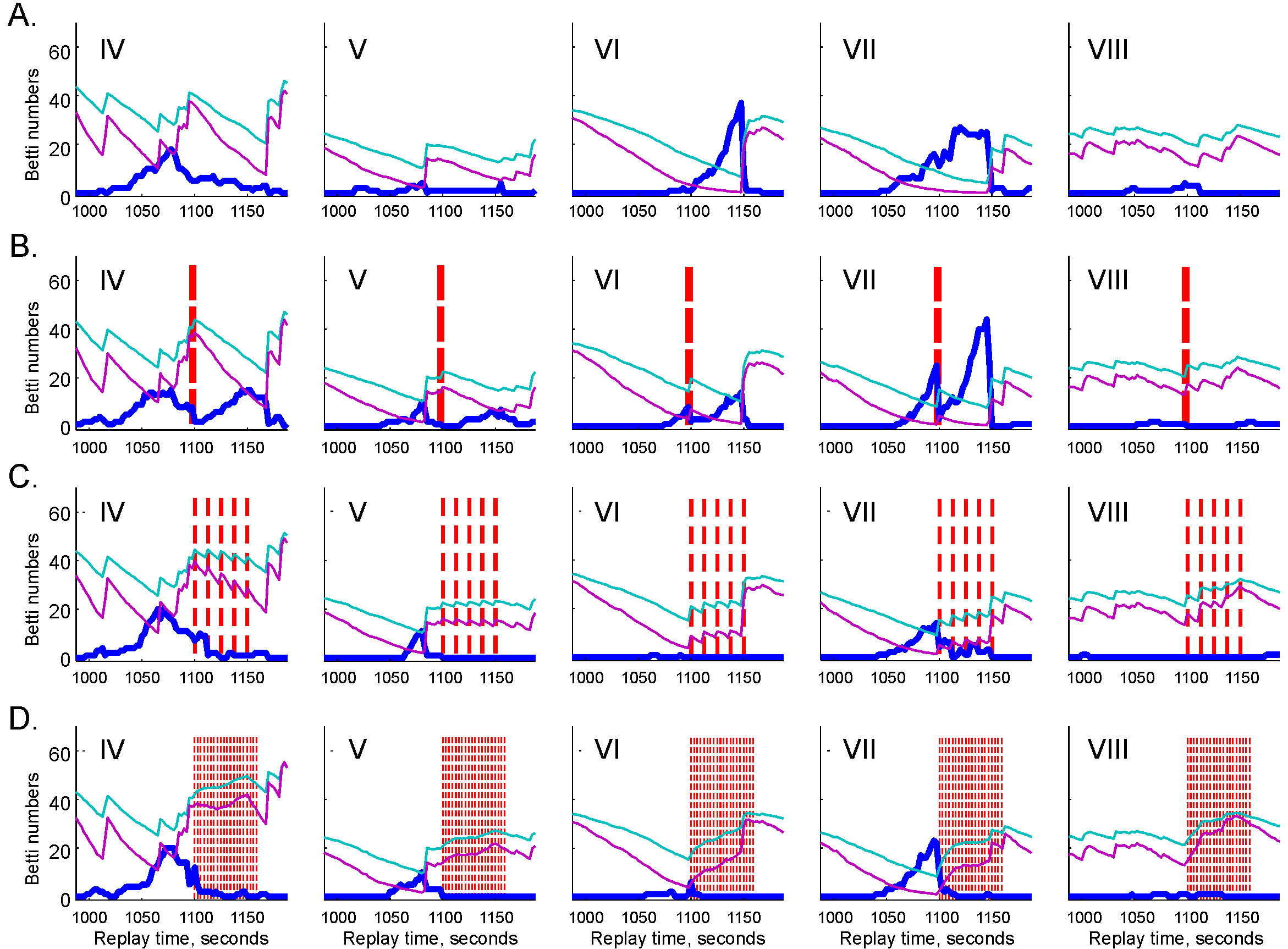}
	\caption{{\footnotesize \textbf{Local topological fluctuations.} \textbf{A}. The first Betti number, $b_1$ (thick blue line) computed for 
			the sectors IV-VIII of the environment, enumerated according to Fig.~\ref{Figure2}D. The changes in numbers of cliques and in Betti 
			numbers over the segments I, II and III are insignificant and are not shown. The blue and the magenta lines represent the numbers 
			of pair and triple connections, $N_2(\mathcal{F}_{\tau})$ and $N_3(\mathcal{F}_{\tau})$, scaled down by the factors $10^{-3}$ and 
			$10^{-4}$ respectively, to fit into the panels. Prior to the instability period, the numbers of $N_3(\mathcal{F}_{\tau})$ and 
			$N_2(\mathcal{F}_{\tau})$ decrease, indicating that the coactivity complex $\mathcal{F}_{\tau}$ thins out, which also produces an 
			increasing number of spurious topological loops. In contrast, the increase of the local Betti numbers $b_1$ stops when the 
			connections start to accumulate, i.e., when the decay of the coactivity simplexes is counterbalanced by their regular reactivation 
			due to the rat's uniform sampling of the environment. 
			\textbf{B}. In the case of instantaneous massive replay (memory flash, time marked by the vertical red dashed line), the increase 
			of $b_1$ is briefly halted over all segments of the environment, but then it restarts at the same rate. 
			\textbf{C}. For a more frequent replays ($N_r = 5$, times marked by the five red vertical dashed lines), spurious loops are suppressed 
			in all sectors, most effectively over the sectors VI and VII. 
			\textbf{D}. In the case of a more regular replay ($N_r =20$ vertical dashed lines), spurious loops are effectively terminated in all 
			sectors over the entire instability period.}}
	\label{SFigure2}
\end{figure}

\begin{figure}[ht]
	\includegraphics[scale=0.83]{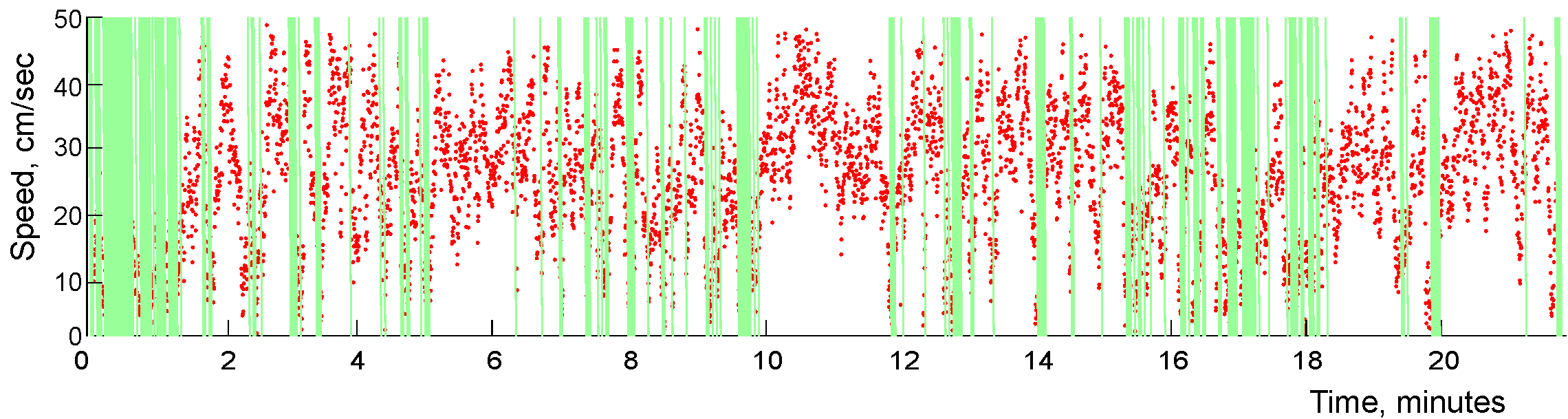}
	\caption{{\footnotesize \textbf{Speed modulation of the replays.} The red dots mark the mean values of the animal's speed during the 
			coactivity windows---the potential replay times. The green lines mark the slow motion periods, $v < 15$ cm/sec, occurring during 
			$14\%$ of time---the replay windows.}}
	\label{SFigure3}
\end{figure} 

\clearpage

\section*{SUPPLEMENTARY MOVIE CAPTIONS}
\label{section:SupplMovies}

\textbf{Suppl. Movie 1.}
\textbf{Single replay.}
First two panels show the dynamics of the spatial histograms of the two-vertex and the three-vertex simplexes (i.e., centers of the pairwise 
and the triple overlaps between place fields) present in $\mathcal{F}_{\tau}$. The right panel shows the trajectory (green line) in the square
environment split into eight segments (see Fig.~\ref{Figure2}C,D). The numbers in each segment $S_i$ represent the pair local Betti numbers 
$(b_0(S_i), b_1(S_i))$ and the pair of the global Betti number $(b_0(\mathcal{E}), b_1(\mathcal{E}))$ are shown in the center. Around $t = 24$ 
secs, a large number of replayed sequences is injected (see Fig.~\ref{Figure3}B). The topological fluctuations are instantaneously suppressed 
but then they immediately restart and reach back to high values. 

\textbf{Suppl. Movie 2.}
\textbf{Five replays} suppress the topological fluctuations better (Fig.~\ref{Figure3}C).

\textbf{Suppl. Movie 3.}
\textbf{Twenty replays} (about one replay in every 9 seconds) nearly extinguish the topological fluctuations during the instability period 
(Fig.~\ref{Figure3}D).

\textbf{Suppl. Movie 4.}
\textbf{Additional speed-modulation} revives topological fluctuations (Fig.~\ref{Figure4}C,D).

\textbf{Suppl. Movie 5.}
\textbf{Speed-modulated, randomized replays} restore the correct topological shape of the map (Fig.~\ref{Figure4}F).
\end{document}